\documentclass[useAMS]{gGAF2e}
\def\etal{\mbox{\it et al.}}

\usepackage{graphicx}
\usepackage{color}
\usepackage{multiequations}

\begin{document}

\doi{10.1080/0309192YYxxxxxxxx}
 \issn{1029-0419}
\issnp{0309-1929}
\jvol{00} \jnum{00} \jyear{2009} \jmonth{February}

\markboth{T. Neukirch}{Geophysical and Astrophysical Fluid Dynamics}
\title{Three-dimensional analytical magnetohydrostatic equilibria of rigidly rotating magnetospheres in cylindrical geometry}

		 \author{ Thomas Neukirch$^{\ast}$\thanks{$^\ast$ Email: thomas@mcs.st-and.ac.uk\vspace{6pt}}\\\vspace{6pt}
		 {\em{School of Mathematics and Statistics,
         University of St. Andrews,
	         St. Andrews KY16 9SS, UK}}\\\vspace{6pt}\received{date}}

\maketitle

\begin{abstract}
We present  three-dimensional solutions of the magnetohydrostatic equations in the co-rotating frame of
reference outside a magnetized rigidly rotating 
cylinder. 
We make no symmetry assumption for the magnetic field, but to be able to make analytical progress
we neglect outflows and specify a particular form for the current density. The magnetohydrostatic equations can then be reduced to a single linear partial
differential equation for a pseudo-potential $U$, from which the magnetic field can be calculated by differentiation. The equation for $U$ can be solved by
standard methods. The solutions can also be used to determine the plasma pressure, 
density and temperature as functions of all three spatial 
coordinates. Despite the obvious limitations of this approach, it can
for example be used as a simple tool to create 
three-dimensional models for the closed field line regions of rotating magnetospheres without rotational symmetry.
\end{abstract}

\begin{keywords}magnetohydrodynamics; analytical solutions; rotating magnetospheres; three-dimensional equilibria 
\end{keywords}\bigskip

\section{Introduction}

Three-dimensional analytical solutions of the magnetohydrostatic equations are difficult to find and only  few solutions are known explicitly. A systematic method for finding a special class of three-dimensional magnetohydrostatic equilibria has been developed by in a series of papers by Low (\citeyear{low85}, \citeyear{low91}, \citeyear{low92}, \citeyear{low93a}, \citeyear{low93b}) and Bogdan and Low (\citeyear{bogdan:low86}). The method relies on the presence of an external force derived from a potential, for example gravitation, and assumes that  the electric current density has specific properties to allow analytical progress. In the simplest possible case, the fundamental equation to be solved is linear and it can be shown that in Cartesian and spherical geometry with external gravity it is very similar to a Schr\"odinger equation \citep{neukirch95a,neukirch:rastatter99}. Therefore, in the linear case, solutions can be found using standard methods, like expansion in terms of orthogonal function systems \citep{rudenko01} or Green's functions \citep{petrie:neukirch00}. Some solutions have also been found for nonlinear cases \citep{neukirch97b}. 
Solutions found by this method have, for example, been used for models of solar and stellar coronae 
\citep[e.g.][]{zhao:hoeksema93,zhao:hoeksema94,gibson:bagenal95,gibson:etal96,zhao:etal00,ruan:etal08, lanza08}.
Another, but  less general method for finding three-dimensional magnetohysdrostatic equilibria in external gravitational fields has been proposed by Osherovich (\citeyear{osherovich85a}, \citeyear{osherovich85b}).

So far the method has only been used to find three-dimensional solutions of the magnetohydrostatic equations in Cartesian or spherical geometry with external gravity, although \citet{low91} also generalizes the method to more general external potentials and forces, including in particular the centrifugal force in a rigidly rotating system. So far, no three-dimensional solutions to the magnetohydrostatic equations have been found for this case and it is the aim of this paper to close this gap. Such solutions could, for example, be used to model the closed field line regions of rotating magnetospheres without symmetries, such as those of the outer planets. 

We restrict our analysis
to the somewhat artificial case of magnetic fields outside a rigidly rotating cylinder, but purely for reasons of mathematical convenience as it is much easier to deal with the boundary conditions on a cylindrical boundary.  It is in principle also possible to treat the more realistic case of a 
spherical body, but imposing boundary conditions would be more difficult.

Planetary magnetospheres are embedded in the solar wind  (or a stellar wind in the case of exoplanets) and the boundary between the closed field line regions and the solar wind/open field line regions is a free boundary determined by total pressure balance, which would have to be calculated as part of the solution. The same statement applies to modeling the closed field line region of a stellar magnetosphere, where the boundary between closed and  open field line regions carrying a stellar wind would have to be determined as part of the solution (in a first idealized step one could maybe try to model the open field line regions by a potential field).  There is little hope that any analytical progress could be made if the calculation of these free boundaries is included in the problem, as finding three-dimensional analytic solutions of the magnetohydrostatic equations is a formidable problem even without considering free boundaries.  The solutions presented in this paper should therefore only be considered as a step towards solving the global problem of modeling magnetospheres, but not as the full solution.

The paper is organised as follows. In section \ref{sec:theory} we present the basic theoretical framework used in this paper, in section \ref{sec:solutions} we present a number of analytical three-dimensional solutions of the magnetohydrostatic equations in cylindrical geometry and we present a summary and conclusions in section \ref{sec:summary}.

\section{Basic theory}
\label{sec:theory}

We consider a cylinder of radius $R$ and infinite length 
rotating rigidly with constant angular 
velocity $\Omega$ about its symmetry axis, which we take to be identical with the $z$-axis.
We will use a cylindrical coordinate system 
$\varpi$, $\phi$, $z$ in a frame of reference co-rotating with the cylinder.

In this frame of reference the MHD equations  are
\citep[see e.g.][chapter 5]{mestel99}
\begin{multiequations}
\singleequation
\begin{eqnarray}
\mathbf{j} \times \mathbf{B} -\nabla p -\rho \nabla V & = & \mathbf{0}, \label{forcebal} \\
\nabla \times \mathbf{B} & = &  \mu_0 \mathbf{j} , \label{ampere} \\
\nabla \cdot \mathbf{B} & = & 0 , \label{divb0} 
\end{eqnarray}
\end{multiequations}
where
$\mathbf{B}$ is the magnetic field, 
$\mathbf{ j}$ is the current density,
$p$       is the pressure,
$\rho$   is the plasma density and
\begin{equation}
V = - \frac{1}{2} \Omega^2 \varpi^2 
\label{vpot}
\end{equation}
is the centrifugal 
potential.

Following \citet{low91}, we
assume that
\begin{equation}
\mu_0\mathbf{ j} = \nabla F \times \nabla V.
\label{jdef}
\end{equation}
Substituting this into the force balance equation (\ref{forcebal}) 
results in
\begin{equation}
\frac{1}{\mu_0} (\mathbf{B} \cdot  \nabla F) \nabla V -
\frac{1}{\mu_0}(\mathbf{B} \cdot \nabla V ) \nabla F  -\nabla p - \rho \nabla V 
= \mathbf{0},
\label{forcebal_low}
\end{equation}
which implies that
\begin{equation}
p(\varpi,\phi,z) = p(F,V).
\end{equation}
It follows that
\begin{multiequations}
\singleequation
\begin{eqnarray}
\left(\frac{\partial p}{\partial F}\right)_V & = &
                     -\frac{1}{\mu_0}(\mathbf{B} \cdot \nabla V ) \label{pequation}\\
\rho &= & - \left(\frac{\partial p}{\partial V}\right)_F +
             \frac{1}{\mu_0} (\mathbf{B} \cdot  \nabla F)       . 
\label{rhoequation}
\end{eqnarray}
\end{multiequations}
The possibility of further analytical progress now depends on the choice of the function $F$. \citet{low91}
suggests the choice 
\begin{equation}
F(\varpi, \phi,z) = \kappa(V)\mathbf{B}\cdot\nabla V,
\label{Fequation}
\end{equation}
where $\kappa(V)$ is a free function of the theory.
The current density is then given by
\begin{equation}
\mu_0\mathbf{j}= \kappa(V)\nabla (\mathbf{B}\cdot\nabla V) \times \nabla V ,
\label{jassumption}
\end{equation} 
and because
the current density depends linearly on the magnetic field,
Amp\`{e}re's law (\ref{ampere}) is also linear in $\mathbf{B}$. Other choices for $F$ are possible, but 
lead to Amp\`{e}re's law being nonlinear in $\mathbf{B}$ 
\citep[for an example in Cartesian geometry see][]{neukirch97b}.

Substituting  (\ref{Fequation}) into  
(\ref{pequation}) we can rewrite it in the form
\begin{equation}
\left(\frac{\partial p}{\partial F}\right)_V  = 
                     -\frac{1}{\mu_0 \kappa(V)} F .
\end{equation} 
Direct integration gives
\begin{equation}
p = p_0(V) -\frac{1}{2\mu_0 \kappa(V)} F^2 .
\label{psol1}
\end{equation}
Substituting  (\ref{Fequation}) for $F$ into  (\ref{psol1}), we finally
obtain
\begin{equation}
p = p_0(V) - \frac{1}{2 \mu_0} \kappa(V) (\mathbf{B}\cdot\nabla V)^2
\label{psol2}
\end{equation}
for the pressure. Here, 
$p_0(V) $ is a positive function which represents a hydrostatic background 
pressure.

The expression for the density can be evaluated by using
 (\ref{psol1}) to get
\begin{equation}
\left(\frac{\partial p}{\partial V}\right)_F =
\frac{{\rm d} p_0}{{\rm d}V} +\frac{1}{2\mu_0 \kappa^2(V)} \frac{d \kappa}{dV} F^2.
\end{equation}
Substituting this into (\ref{rhoequation}) and using (\ref{Fequation})
the final expression for the density finally becomes
\begin{equation}
\rho=-\frac{{\rm d} p_0}{{\rm d}V} + \frac{1}{2\mu_0}\; \frac{{\rm d} \kappa}{{\rm d} V}\; 
(\mathbf{B}\cdot\nabla V)^2 +
+\frac{1}{\mu_0}\;\kappa(V) \mathbf{B}\cdot\nabla (\mathbf{B}\cdot\nabla V).
\label{rhosol1}
\end{equation}
This form of the plasma pressure and density is a direct consequence of the assuming that current density has the form (\ref{jdef}) with the function $F$ given by (\ref{Fequation}).

It can be seen that the term representing the background density, $\rho_0(V)$, is given by 
\begin{equation}
\rho_0 (V) = - \frac{{\rm d} p_0}{{\rm d}V}.
\end{equation}
For the case discussed in the present paper, $V$ is strictly negative and the density is strictly positive, which implies the background pressure must be monotonically increasing with distance from the rotation axis, with the precise form depending on the form of the background density. We shall not present any detailed discussion of a specific model for the hydrostatic background plasma in this paper, but it is important to choose the hydrostatic background plasma in such a way that it guarantees that the full density and pressure are always positive, because that is not automatically guaranteed.
Also the background pressure will become important if one wants to solve the full magnetospheric problem including the free boundaries because continuity of total pressure across the free boundary is a necessary condition for an equilibrium solution.

Although in the present paper we will only make use of the potential (\ref{vpot}) we remark that in deriving the expressions for the pressure and the density,  (\ref{psol2}) and
(\ref{rhosol1}), no use has been made of the particular coordinate system or form of the
potential $V$. It should be emphasized, however, that this does of course not imply that this theory can be generalized to differentially rotating systems as that would violate Ferraro's  iso-rotation theorem \citep{ferraro37}. A permitted generalization of the potential (\ref{vpot}) would be to add a gravitational potential, i.e. instead of having just a magnetized central body one would have a massive magnetized central body. The shape of the body and thus the form of the potential $V$ is irrelevant for the validity of  (\ref{psol2}) and (\ref{rhosol1}), as long as the gravitational potential is time-independent in the rotating frame of reference. 

An expression for the plasma temperature can be obtained if we assume that
the plasma satisfies the equation of state of an ideal gas,
\begin{equation}
T= \frac{\mu p}{ \mbox{R} \rho},
\end{equation}
where $\mbox{R}$ is the universal gas constant and $\mu$ is the mean molecular weight.

The magnetic field can be calculated using Amp\`ere's law (\ref{ampere}).
The current density can be written
as
\begin{equation}
\mu_0\mathbf{j} = \nabla \times (F\nabla V)
\end{equation}
so that
\begin{equation}
\nabla\times \mathbf{B} = \nabla \times (F\nabla V).
\label{ampere2}
\end{equation}
This equation can be integrated with the result
\begin{equation}
\mathbf{B} = \nabla U + F \nabla V,
\end{equation}
where $U$ is a free function. Using  (\ref{Fequation}) to substitute in for $F$, we obtain
\begin{equation}
\mathbf{B} = \nabla U + \kappa(V) (\mathbf{B}\cdot \nabla V) \nabla V .
\label{B1}
\end{equation}
Note that this is not yet the final expression for $\mathbf{B}$ as the
right hand side still depends on $\mathbf{B}$.
Taking the scalar product of 
(\ref{B1}) with  $\nabla V$ 
and solving for $\mathbf{B}\cdot \nabla V$, we obtain
\begin{equation}
\mathbf{B}\cdot \nabla V = \frac{\nabla U\cdot\nabla V }{1- \kappa(V) (\nabla V)^2} .
\end{equation}
The final expression for the magnetic field therefore is
\begin{equation}
\mathbf{B} = \nabla U + 
\frac{\kappa(V) }{1- \kappa(V) (\nabla V)^2} (\nabla U\cdot\nabla V )\nabla V .
\label{B2}
\end{equation}
Again, expression (\ref{B2}) is completely independent of the form of $V$, bearing in mind the constraints discussed before. We also emphasize again that the specific form of  (\ref{B2}) is a direct consequence of  (\ref{jdef}) and (\ref{Fequation}). 

For the form (\ref{vpot}) of the potential we find the components of $\mathbf{B}$ to be
\begin{multiequations}
\singleequation
\begin{eqnarray}
B_\varpi&  = & \frac{1}{1-\kappa(V) (V^\prime)^2} \frac{\partial U}{\partial \varpi}, \label{Br2}   \\
B_\phi  &  = & \frac{1}{\varpi}\frac{\partial U}{\partial \phi},                      \label{Bphi2} \\
B_z     &  = & \frac{\partial U}{\partial z}                                       \label{Bz2}
\end{eqnarray}
\end{multiequations}
with
\begin{equation}
V^\prime = \frac{{\rm d}V}{{\rm d}\varpi} .
\end{equation}
From  (\ref{B2}) and the equations for the components of the
magnetic field, (\ref{Br2}), (\ref{Bphi2}) and (\ref{Bz2}), we see that
$U$ is a pseudo-potential, from which $\mathbf{B}$ can be determined by differentiation. 
We call it a pseudo-potential because
the expression for $B_\varpi$ is modified compared to that derived from a normal potential, where $\mathbf{B}$ would be given by the
the gradient of the potential.
This modification is due to the presence of currents in the
system, represented by the factor $1/(1-\kappa {V^\prime}^2)$ in the equation for 
$B_\varpi$,  (\ref{Br2}). 

Finally, the pseudo-potential $U$ is determined by substituting  (\ref{B2})
into the solenoidal condition  (\ref{divb0}), giving
\begin{equation}
\nabla \cdot \left(
\nabla U + 
\frac{\kappa(V) }{1- \kappa(V) (\nabla V)^2} (\nabla U\cdot\nabla V )\nabla V
\right) =0 .
\label{UeqV}
\end{equation}
Again, this equation is valid for general forms of the potential $V$, and not just for the form used in the present paper. An alternative form of this equation is
\begin{equation}
\nabla \cdot ({\textrm{\bf M}}\cdot\nabla U) = 0,
\label{UeqV_alt}
\end{equation}
with the $3\times 3$ matrix ${\textrm{\bf M}}$ defined as
\begin{equation}
{\textrm{\bf M}} = {\textrm{\bf I}} +\frac{\kappa(V) }{1- \kappa(V) (\nabla V)^2}  \nabla V\, \nabla V.
\end{equation}
Here ${\textrm{\bf I}}$ is the $3\times 3$ unit matrix. Equation (\ref{UeqV_alt}) is particularly 
useful if $\nabla V$ has more than one non-vanishing component. This would, for example, be the case if we would treat the present problem in spherical polar coordinates instead of cylindrical coordinates.

For the centrifugal potential  (\ref{vpot}) we define
\begin{equation}
\xi(\varpi) = \Omega^2\varpi^2 \kappa(V)  ,
\label{xidef}
\end{equation}
and rewrite  (\ref{UeqV}) as
\begin{equation}
      \frac{1}{\varpi}\frac{\partial}{\partial \varpi}
\left( \frac{\varpi}{1-\xi(\varpi)} \frac{\partial U}{\partial \varpi} \right) +
\frac{1}{\varpi^2}\frac{\partial^2 U}{\partial \phi^2} +
\frac{\partial^2 U}{\partial z^2} = 0.
\label{Uequation}
\end{equation}

Equation (\ref{Uequation}) is the fundamental equation that has to be solved to calculate 
the three-dimensional magnetic field.
The functions $\kappa(V)$, or alternatively $\xi(\varpi)$ are free functions and can be chosen
to make analytical progress possible.  We remark that it is only possible to choose 
directly the function $\xi(\varpi)$ instead of $\kappa(V)$ because of the one-to-one mapping
between $\varpi$ and $V$ defined by  (\ref{vpot}). In cases where this mapping is not one-to-one, more care has to be taken.

Using the definition (\ref{xidef}) of $\xi(\omega)$ the expressions for the pressure and the density can be rewritten as
\begin{multiequations}
\singleequation
\begin{eqnarray}
p & =&   \bar{p} _0(\varpi) -\frac{1}{2\mu_0} \frac{\xi(\varpi)}{(1-\xi(\varpi))^2}\left(\frac{\partial U}{\partial \varpi}\right)^2      \label{p_xi}  ,   \\
\rho &=&\frac{1}{\Omega \varpi}\left[ \frac{d\bar{p}_0}{d\varpi} - 
\frac{1}{2\mu_0} \frac{1}{(1-\xi(\varpi))^2} \frac{d\xi}{d \varpi}\left(\frac{\partial U}{\partial \varpi}\right)^2    
-\frac{1}{\mu_0}\xi (\mathbf{B}\cdot\nabla) B_\varpi      \right] . \label{rho_xi}
\end{eqnarray}
\end{multiequations}
An important point to note is that  (\ref{p_xi}) and (\ref{rho_xi}) (or alternatively  (\ref{psol2}) and (\ref{rhosol1})) do not automatically guarantee that pressure and density are always positive, but that one can always arrange them to be positive by choosing an appropriate background plasma model.

\section{Example Solutions}
\label{sec:solutions}

Equation (\ref{Uequation}) is very similar to Laplace's equation and one can immediately see
that it has separable solutions of the form
\begin{equation}
U(\varpi,\phi,z) = F_{mk}(\varpi) \exp(\mbox{i} m \phi) \exp(\mbox{i} k z) .
\label{Usolsep}
\end{equation}
with $m$ integer and $k$ real.
If we substitute  (\ref{Usolsep}) into  (\ref{Uequation}) we find that
the radial function $F_{mk}(\varpi)$ satisfies the equation
\begin{equation}
\frac{1}{\varpi}\frac{{\rm d}}{{\rm d} \varpi}
\left( \frac{\varpi}{1-\xi(\varpi)} \frac{{\rm d}F_{mk}}{{\rm d} \varpi} \right) - 
\left( \frac{m^2}{\varpi^2} + k^2\right) F_{mk} = 0
\label{radialequation}
\end {equation}
This ordinary second order differential equation will generally have two linearly independent solutions,
$F^{(1)}_{mk}(\varpi)$ and $F^{(2)}_{mk}(\varpi)$, say.
Since the partial differential equation for $U$ is linear
the solutions for different $m$ and $k$ may be superposed to give the general solution in the form
\begin{equation}
U(\varpi,\phi,z) = \sum_{m=-\infty}^\infty  \exp(\mbox{i} m \phi) 
                   \int_{-\infty}^\infty {\rm d}k\,  \exp(\mbox{i} k z) [A_m(k) F_{mk}^{(1)}(\varpi) +
                                             B_m(k) F_{mk}^{(2)}(\varpi)].
 \label{Usolution}                                                
\end{equation}
The complex coefficients $A_m(k)$ and $B_m(k)$  
will be determined by the boundary conditions imposed on
the surface of the cylinder ($\varpi=R$) and on the outer boundary
($\varpi=\varpi_{out}$) and can be calculated using the standard formulae from the theory of Fourier transformations and series.  From a mathematical point of view, the outer boundary 
could be at infinity, but from a physical point of view the problem will usually
only be considered on a finite domain
since the 
assumptions made for deriving our theory will break down at a finite distance from the rotation axis.

For general choices of $\kappa(V)$, or alternatively, $\xi(\varpi)$, 
 (\ref{Uequation}), or the corresponding radial equation (\ref{radialequation}) 
still has to be solved numerically.
For some cases, however, exact solutions can be found, and we will
discuss a couple of these cases in more detail now.  In the following we assume that $\varpi$ and $z$ have been normalised
by the radius of the cylinder $R$, i.e. that the cylinder radius is $1$ in
these normalised coordinates.

The simplest possible choice for $\xi(\varpi)$ is
\begin{equation}
\xi(\varpi) = \xi_0 = \mbox{constant} .
\end{equation}
In this case Equation (\ref{radialequation}) takes the form
\begin{equation}
\frac{1}{\varpi}\frac{{\rm d}}{{\rm d} \varpi}
\left( \frac{\varpi}{1-\xi_0} \frac{{\rm d} F_{mk}}{{\rm d} \varpi} \right) - 
\left( \frac{m^2}{\varpi^2} + k^2\right) F_{mk} = 0
\label{radialex1}
\end {equation}
with the two linearly independent 
solutions 
\begin{multiequations}
\singleequation
\begin{eqnarray}
F_{mk}^{(1)}(\varpi) &= & I_\nu(k\sqrt{1-\xi_0} \;\varpi), \\       
F_{mk}^{(2)}(\varpi) &= & K_\nu(k\sqrt{1-\xi_0} \;\varpi),
\end{eqnarray}
\end{multiequations}
where $I_\nu(x)$ and $K_\nu(x)$ are modified Bessel 
functions \citep{abramowitz:stegun65} with $\nu = m \sqrt{1-\xi_0} $. Here $\xi_0 < 1$ has been assumed.

Another function $\xi(\varpi)$ for which analytical solutions of  
(\ref{radialequation})
can be found is
\begin{equation}
\xi(\varpi) = 1- q \varpi^2, \qquad q = \mbox{constant} > 0.
\end{equation}
In this case  (\ref{radialequation}) has the form
\begin{equation}
\frac{1}{\varpi}\frac{{\rm d}}{{\rm d} \varpi}
\left( \frac{1}{q \varpi} \frac{{\rm d}F_{mk}}{{\rm d} \varpi} \right) - 
\left( \frac{m^2}{\varpi^2} + k^2\right) F_{mk} = 0 .
\label{radialex2}
\end {equation}
With the coordinate transformation
\begin{equation}
w= \frac{1}{2} \varpi^2,
\end{equation}
  (\ref{radialex2}) can be transformed into
\begin{equation}
\frac{{\rm d}^2F_{mk}}{{\rm d} w^2} - 
q\left( \frac{m^2}{2 w} + k^2\right) F_{mk} = 0 .
\label{radialex2trans}
\end {equation}
The solutions of (\ref{radialex2trans}) can be expressed 
in terms of confluent hypergeometric
functions $M(a,b,x)$ and $U(a,b,x)$ as
\citep[][Chapter 13]{abramowitz:stegun65},
\begin{multiequations}
\singleequation
\begin{eqnarray}
F_{mk}^{(1)}(\varpi) &= &  2\sqrt{q k^2} \varpi \exp(-\sqrt{q k^2}\varpi)  
                                            M(1 +m^2/(8k^2),1,2\sqrt{q k^2}\varpi) \\ 
 F_{mk}^{(2)}(\varpi) &= &  2\sqrt{q k^2} \varpi \exp(-\sqrt{q k^2} \varpi)   
                                              U(1 +m^2/(8k^2),1,2\sqrt{q k^2}\varpi) .
\end{eqnarray}
\end{multiequations}

To calculate model magnetospheres one would now have to 
choose appropriate boundary conditions, determine the coefficients 
$A_m(k)$ and $B_m(k)$ and calculate the pseudo-potential $U$ 
using  (\ref{Usolution}). 
From $U$, the magnetic field, the plasma 
pressure and plasma density can be derived using  (\ref{Br2}) to 
(\ref{Bz2}) and (\ref{p_xi}) to (\ref{rho_xi}). This will usually involve 
expressions containing an infinite series and an integral, which, although 
not inherently difficult, can make the algebra a bit tedious, especially 
when calculating the pressure and density as terms that are quadratic in the magnetic field 
and its derivatives are involved. 

For the sake of simplicity we will use a different method to calculate and discuss 
an illustrative example solution for the case $\xi(\varpi) = \xi_0 <1$. Defining
\begin{multiequations}
\singleequation
\begin{eqnarray}
\bar{\varpi} &=&\sqrt{1-\xi_0} \varpi , \label{varpi_scaled}\\
\bar{\phi}    &=& \frac{1}{\sqrt{1-\xi_0}} \phi  , \label{phi_scaled}
\end{eqnarray}
\end{multiequations}
 (\ref{Uequation}) becomes
\begin{equation}
      \frac{1}{\bar{\varpi}}\frac{\partial}{\partial\bar{ \varpi}}
\left( \bar{\varpi} \frac{\partial U}{\partial \bar{\varpi}} \right) +
\frac{1}{\bar{\varpi}^2}\frac{\partial^2 U}{\partial \bar{\phi}^2} +
\frac{\partial^2 U}{\partial z^2} = 0,
\label{Uequation_scaled}
\end{equation}
which is the Laplace equation in cylindrical coordinates $\bar{\varpi}$, $\bar{\phi}$ and $z$. This shows that by rescaling of the coordinates $\bar{\varpi}$ and $\bar{\phi}$ solutions to the Laplace equation (\ref{Uequation_scaled}) can be transformed into solutions of  (\ref{Uequation}). 
A similar method has been used by \citet{petrie:neukirch00} for finding a closed expression for
a Green's function for the constant $\xi$-case in Cartesian coordinates.
The only 
constraint is that the transformed solutions have to be $2\pi$-periodic in the $\phi$-coordinate. 
It should be noted that this is not a requirement for the solutions of the transformed Laplace 
equation which could be used to satisfy the periodicity condition in the untransformed 
coordinate $\phi$. Another important point is 
that we are of course not restricted to using solutions to the Laplace equation found in cylindrical 
coordinates, instead any other coordinate system could be used, e.g.  spherical coordinates. We could even allow an inner boundary with a shape different from a cylinder, e.g. a sphere. The major disadvantage of this solution method compared to the standard separation of variables method is that it will be impossible to match given boundary conditions. Also, the method only works for the $\xi(\varpi) =$ constant case, but not for other choices of $\xi(\varpi)$.

Despite these limitations we shall now use the transformation method to generate and discuss an illustrative solution of the general method described in this paper. As we are only interested in presenting an example solution we shall simply assume boundary conditions which match the solution we arrive at by using the transformation. As a starting point we shall use a magnetic dipole field but we allow the position of the dipole to be off-axis and the direction of the dipole to be non-aligned with the symmetry axis of the cylinder. Without loss of generality we can assume that the position of the dipole is $\bar{\varpi} =\bar{\varpi}_d$, $\bar{\phi}=0$ and $z=0$. In the corresponding Cartesian coordinate system, the dipole sits on the $x$-axis at a distance $\bar{\varpi}_d$ from the origin.

The appropriate solution of the Laplace equation is
\begin{equation}
U(\bar{\varpi},\bar{\phi},z) = - \frac{\mu_0 m}{4\pi} \frac{
\sin\Theta \cos\Psi(\bar{\varpi} \cos\bar{\phi} -\bar{\varpi}_d)+\sin\Theta
                                                                                                    \sin\Psi \bar{\varpi}\sin\bar{\phi}
                                                                                                  + \cos\Theta  z }
       {(\bar{\varpi}^2 -2 \bar{\varpi}_d \bar{\varpi}\cos\bar{\phi} +\bar{\varpi}_d^2 +z^2)^{3/2}},
\label{Udipole}
\end{equation}
where $m$ is the magnitude of the magnetic dipole moment, $\Psi$ is the angle between the direction of the dipole moment and the direction of the $\bar{\varpi}$-axis and $\Theta$ the angle between the direction of the dipole moment and the $z$-axis.

In order to make the transformed solution $2\pi$-periodic, the argument of the trigonometric 
functions in $U$ has to be an integer multiple of $\phi$. This can be satisfied by letting, for example,
\begin{multiequations}
\singleequation
\begin{eqnarray}
\xi_0  &  = & \frac{3}{4}  ,               \\
\sqrt{1-\xi_0} & = & \frac{1}{2}, \\
\bar{\phi} & =& 2\phi.
\end{eqnarray} 
\end{multiequations}
Other possible choices for the value of $\xi_0$ are $\xi_0= 1 - n^{-2}$, $n > 2$ integer, 
which gives $\bar{\phi} = n\phi$.

Normalising $\varpi$ and $z$ by the cylinder radius $R$ and the potential $U$ 
by $\mu_0 m/(\pi R^2)$, we obtain
\begin{equation}
U(\varpi,\phi,z) =-\frac{
\sin\Theta \cos\Psi(\varpi \cos(2\phi)-\varpi_d)+\sin\Theta
                                                                                                    \sin\Psi \varpi \sin(2\phi)
                                                                                                  + 2\cos\Theta  z }
       {(\varpi^2 -2 \varpi_d \varpi\cos(2\phi) +\varpi_d^2 + 4 z^2)^{3/2}},
\label{Udipole_scaled}
\end{equation}
where $\varpi_d = 2\bar{\varpi}_d$.

\begin{figure}[t]
\begin{center}
\includegraphics[width=0.3\textwidth]{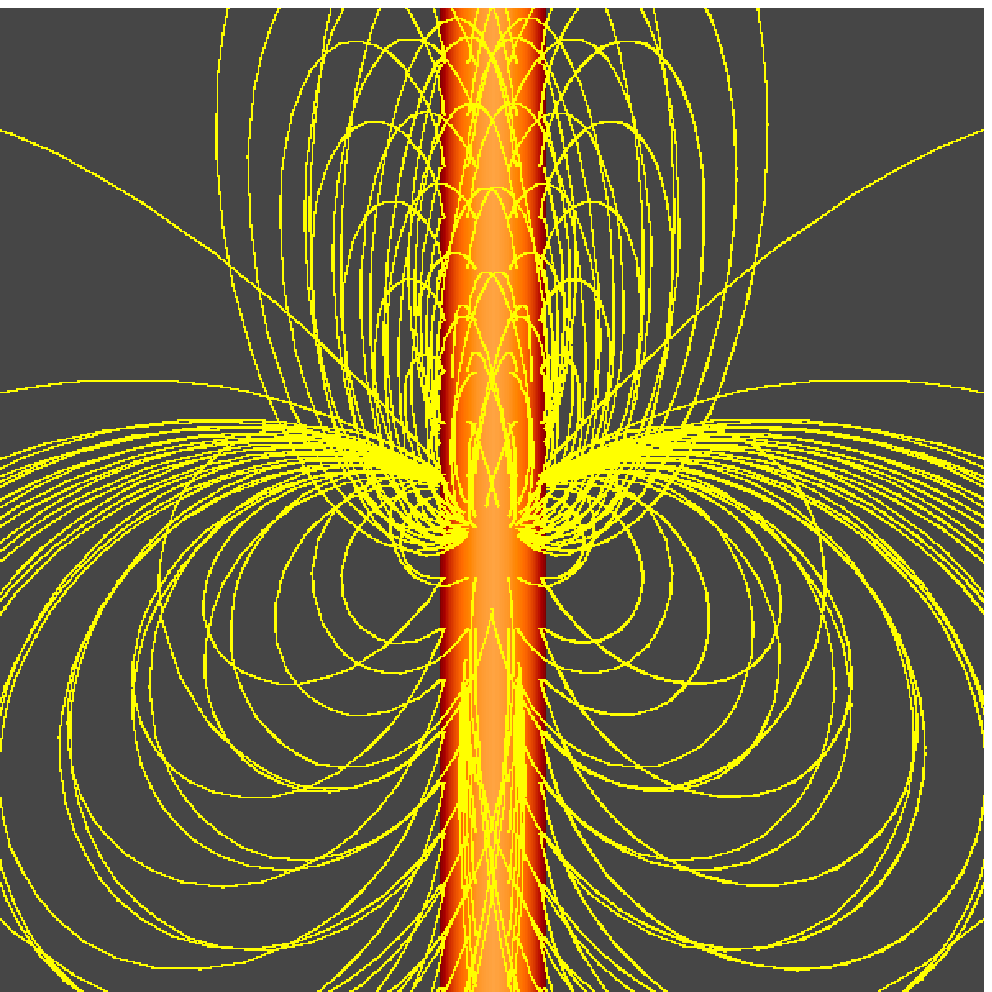}
\includegraphics[width=0.3\textwidth]{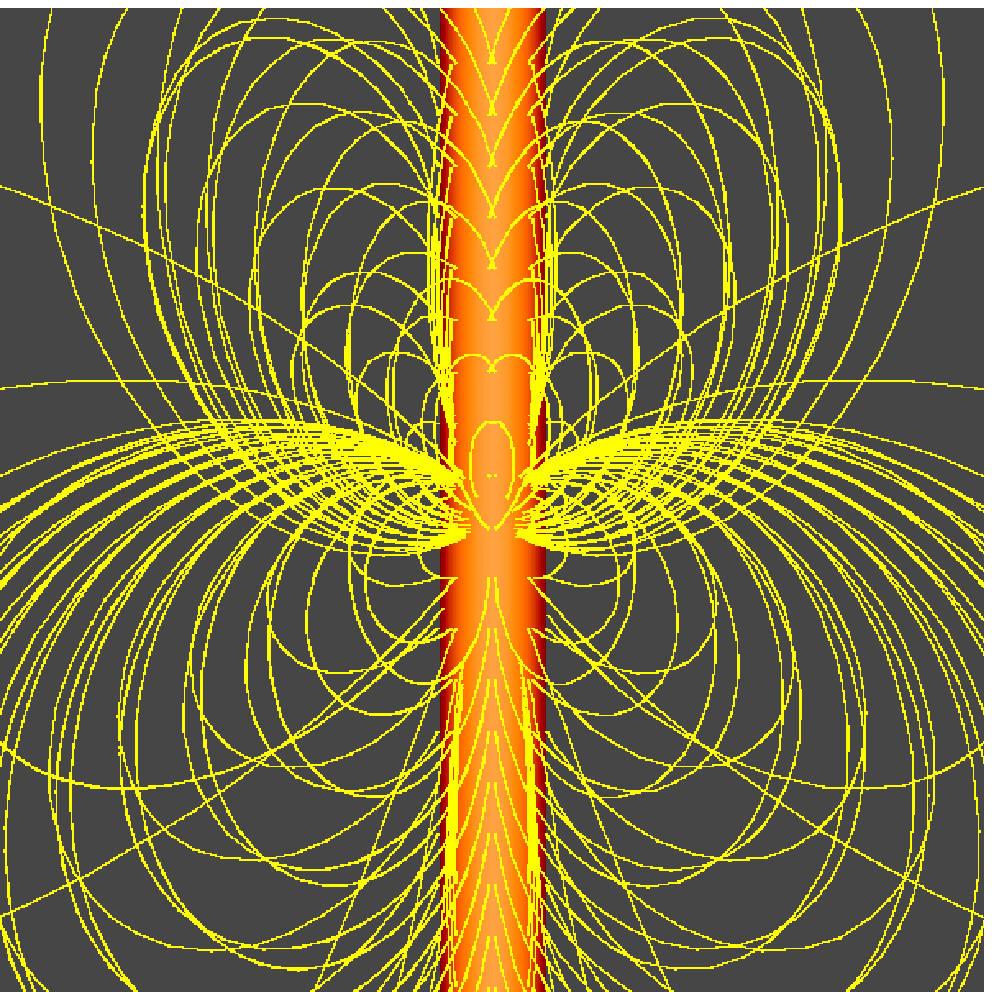}
\includegraphics[width=0.3\textwidth]{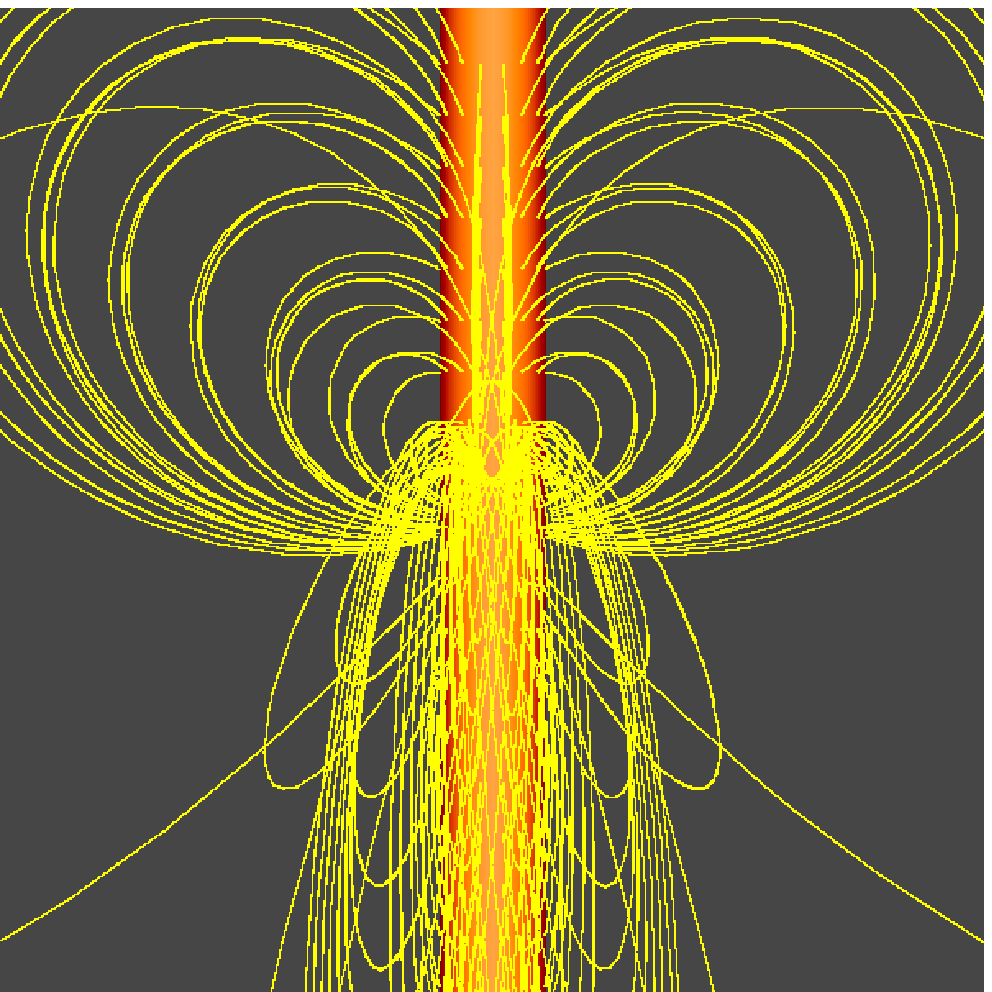}
\vfil
\end{center}
\caption{Three different viewing angles ($\phi=0$, $\phi=2\pi/3$ and $\phi=4\pi/3$) of a 3D plot of selected field lines for the
example solution for $\xi(\varpi)=\xi_0=3/4$ and the dipole parameters $\varpi_d =0.5$, $\Theta = \pi/4$
and $\Psi =\pi/4$.}
\label{fig:exampleB}
\end{figure}

Using the abbreviations
\begin{multiequations}
\singleequation
\begin{eqnarray}
D &=& \sin\Theta \cos\Psi(\varpi \cos(2\phi)-\varpi_d)+\sin\Theta
                                                                                                    \sin\Psi \varpi \sin(2\phi)
                                                                                                  + 2\cos\Theta  z, \label{denom_u}  \\
r & = &   (\varpi^2 -2 \varpi_d \varpi\cos(2\phi) +\varpi_d^2 + 4 z^2)^{1/2} ,   \label{nom_u}                                                                                           
\end{eqnarray}
\end{multiequations}
the magnetic field components are given by
\begin{multiequations}
\singleequation
\begin{eqnarray}
B_\varpi & = &  \frac{12D(\varpi-\varpi_d \cos(2\phi)) - 
4\sin\Theta(\cos\Psi\cos(2\phi)+\sin\Psi\sin(2\phi))r^2}
{r^5} ,
\label{Br_example}\\
%
%
B_\phi    & = & \frac{6 D\varpi_d  \sin(2\phi)
-2\sin\Theta(\sin\Psi \cos(2\phi) -\cos\Psi\sin(2\phi))r^2}
{r^5} ,
\label{Bphi_example} \\
%
%
B_z        & = &   \frac{12Dz-2\cos\Theta  r^2}{r^5} .
\label{Bz_example}
%
%
\end{eqnarray}
\end{multiequations}
The normalising factor for the magnetic field is $B_0 = (\mu_0 m)/ (\pi R^3)$.
In Figure \ref{fig:exampleB} we show the magnetic field for the case $\varpi_d=0.5 R$, 
$\Theta= \pi/4$ and $\Psi = \pi/4$. The same set of field lines is shown from three different perspectives corresponding to the viewing angles $\phi=0$, $2\pi/3$ and $4\pi/3$.

\begin{figure}[t]
\begin{center}
\includegraphics[width=0.45\textwidth]{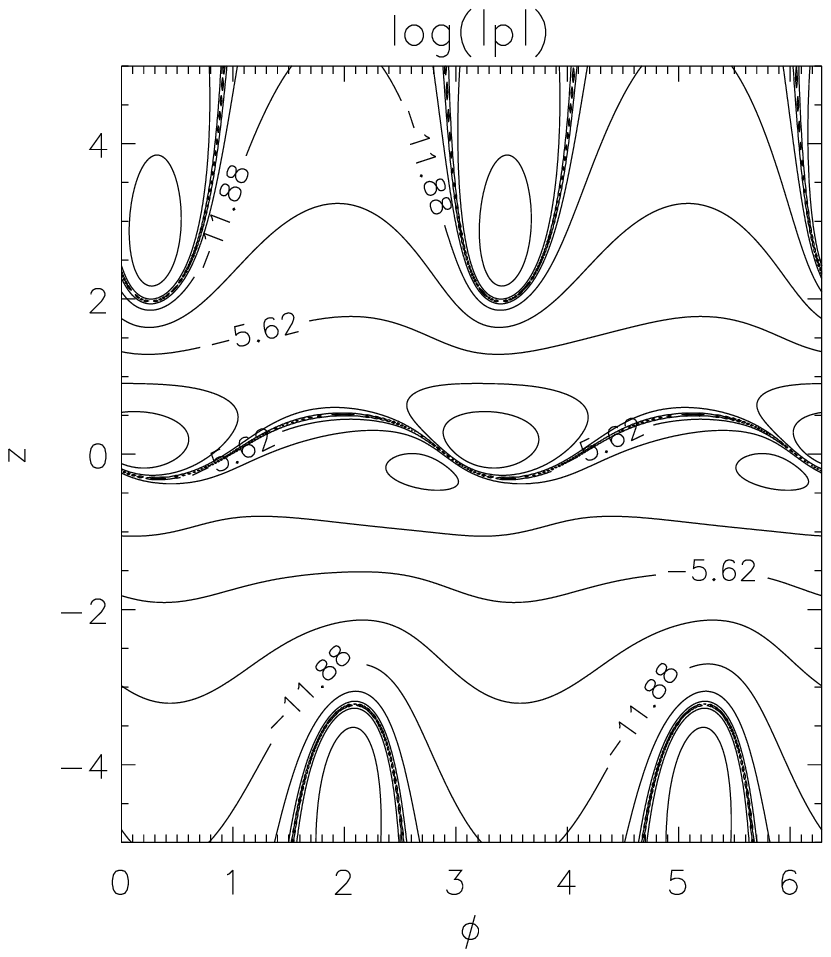}
\includegraphics[width=0.45\textwidth]{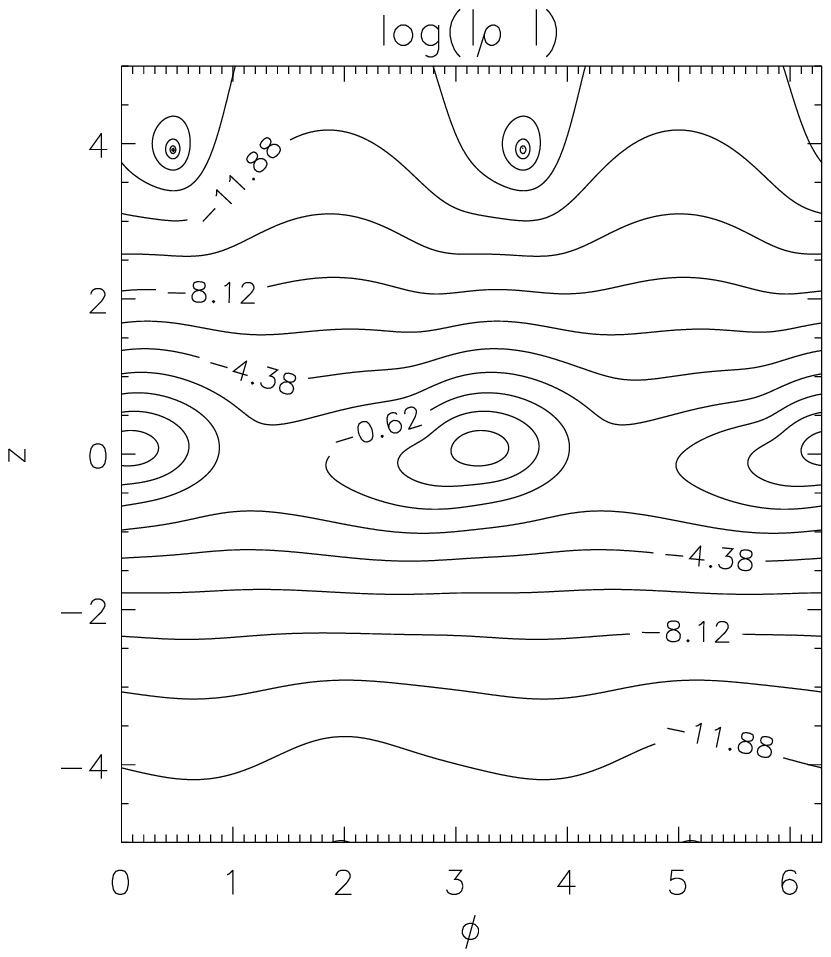}

\includegraphics[width=0.45\textwidth]{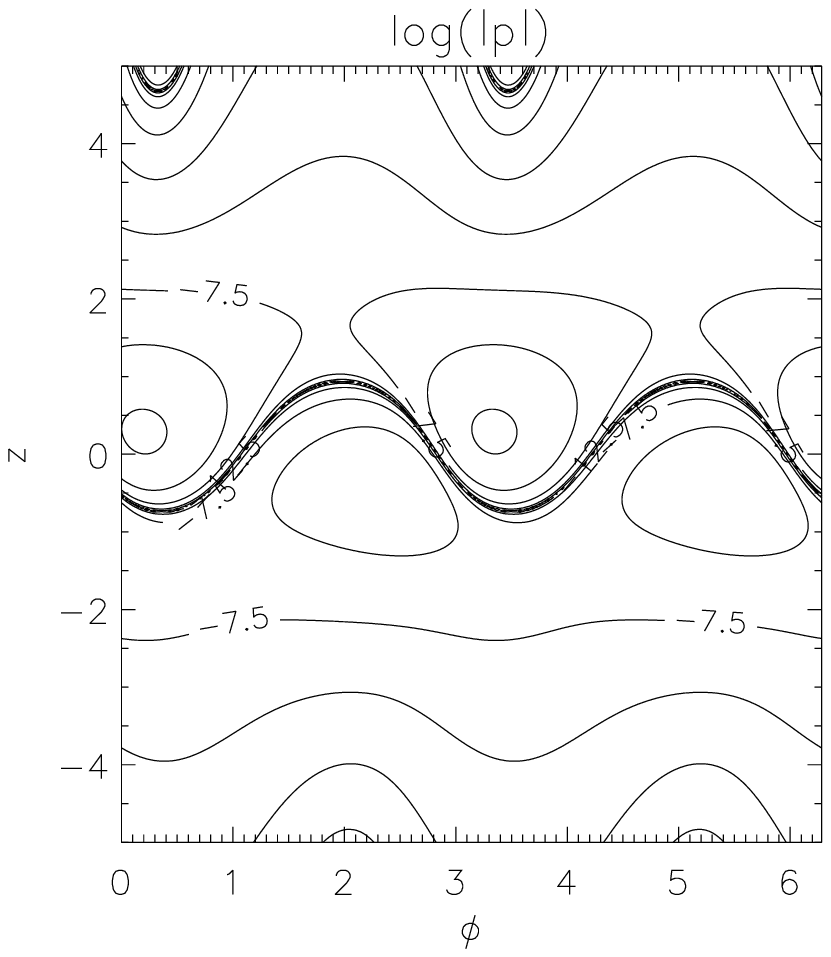}
\includegraphics[width=0.45\textwidth]{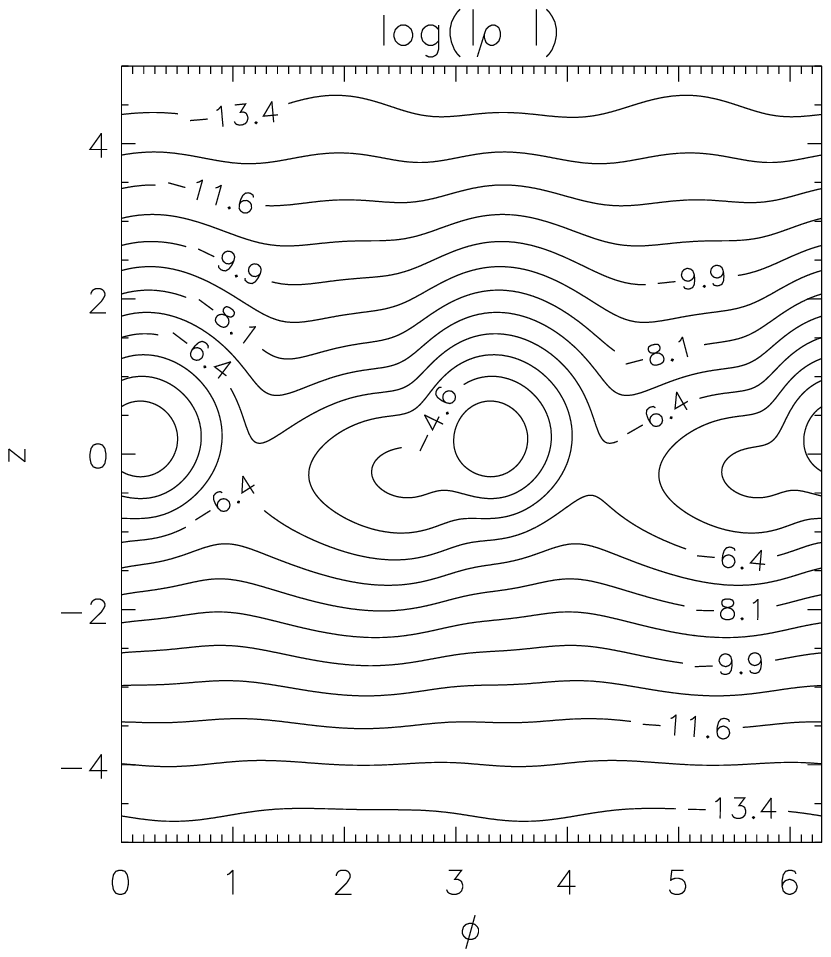}
\vfil
\end{center}
\caption{Contour plots in the $\phi$-$z$ plane at radii $\varpi=1.5$ (top row) and $\varpi=3.0$ (bottom row) of the natural logarithm of the pressure and the density variation. Note that the background quantities have been subtracted as they can be chosen independently and only the three-dimensional terms are shown.}
\label{fig:pandrho}
\end{figure}

The pressure (normalised to $B_0^2/(2\mu_0)= (\mu_0 m^2)/(2\pi^2 R^6) $) is given by
\begin{eqnarray}
p(\varpi,\phi,z) &=& \bar{p}_0(\varpi) \nonumber \\
& & \mbox{\hspace{-2cm}}-  12
\frac{[3D(\varpi-\varpi_d \cos(2\phi)) - 
\sin\Theta(\cos\Psi\cos(2\phi)+\sin\Psi\sin(2\phi))r^2]^2}
{r^{10}} ,
\label{pressure_example}
\end{eqnarray}
and the density is given by
\begin{equation}
\rho(\varpi,\phi,z) = \frac{1}{2 \varpi }  \left[ \frac{{\rm d} \bar{p}_0}{{\rm d} \varpi}-
\frac{3}{2} \mathbf{B}\cdot\nabla B_\varpi \right],
\label{density_example}
\end{equation}
with the density normalised by $\mu_0 m^2/(\pi^2 \Omega R^8)$. We have refrained from giving the
explicit expression for the second term in the density as it is quite lengthy, but not really contributing to any better understanding of the problem.
We are still free to specify a hydrostatic background plasma to complete our example. 
In Figure \ref{fig:pandrho}
we show contour plots in the $\phi$-$z$-plane for $\varpi=1.5$ and $\varpi=3.0$ of the logarithm  of the pressure and the density. The background terms have been subtracted to emphasize the variation of pressure and density in $\phi$ and $z$. It turns out that for the two radii shown, the pressure term is always negative and we therefore show contours of the natural logarithm of its modulus, whereas the density term is positive. This means that the plasma pressure will be reduced compared to whatever background pressure is chosen, whereas the plasma density will be enhanced above the background plasma density.
One can also see the strong radial decrease of the density term by comparing the values of the contours at $\varpi=1.5$ and $\varpi=3$. This is consistent with  (\ref{density_example}) which shows that the density term should have a $\varpi^{-8}$-dependence, whereas the pressure term has a $\varpi^{-6}$ variation (see  (\ref{pressure_example})). 
Although the full expressions for the density and pressure will depend on the background model, this will usually mean that the three-dimensional variations of pressure and density will only be of importance close to the central cylinder and decrease fast with distance from the rotation axis ($\varpi$).

\section{Summary and Discussion}
\label{sec:summary}

Following \citet{low91} we have presented a relatively simple analytical approach to calculate
three-dimensional analytical MHD equilibria of
 rigidly
rotating magnetospheres in cylindrical geometry.  
The fundamental equation that has to be solved is a linear second order 
PDE for a pseudo-potential $U$. The theory
contains a free function $\kappa(V)$, or alternatively a function $\xi(\varpi)$,
which can be chosen such that analytical progress is possible. 
Using the standard method of separation of variables, analytical solutions have been found
 for two different choices of $\xi(\varpi)$. For the choice of  $\xi(\varpi)=\xi_0$ constant, one can 
 also find solutions using a coordinate transformation method. We have used this method to find and discuss some illustrative solutions.
 
The solutions found here do not represent a complete model of the closed field line regions of rotating magnetospheres, because such a model would include the calculation of the free boundaries between open and closed field line regions and/or the boundaries between the closed field line regions and the plasma in which the magnetosphere is embedded. Determining the free boundaries generally involves matching the total pressure on both sides of the free boundary and obviously the choice of the hydrostatic background plasma, which is part of the set of solutions presented here will be important. Generally, if one would attempt to calculate a model with free boundaries using the results presented in this paper the separation of variable method is better suited than the transformation method due to the availability of a complete set of functions allowing the expansion of the general solution. The difficulty would be the nonlinear nature of the boundary conditions on and the unknown shape of the free boundary.

Possible modifications and extensions of the theory are to use spherical inner and outer boundaries
or to use a combination of centrifugal and gravitational potential. This could, for example, be used to model the closed field line regions of fast rotating stars \citep[see e.g.][]{ryan:etal05}. In cylindrical geometry the modelling process will be very similar to that presented in the present paper, but with the additional complication that the combined potential
\begin{equation}
V(\varpi) = -\frac{1}{2} \Omega^2 \varpi^2 + 2G M \ln(\varpi/R)
\label{centrplusgrav}
\end{equation}
is not monotonic and thus cannot be mapped one-to-one to $\varpi$, because the combined potential has a maximum at the co-rotation radius $\varpi_c$, where the Kepler velocity equals the 
rotational velocity $\Omega \varpi_c$. A simple combination of $\kappa(V) {V^\prime}^2 $ into a function 
$\xi(\varpi)$ will usually lead to singularities in pressure and density at the co-rotation radius.  Numerical solutions for physically sensible choices of $\kappa(V)$ are, however, possible in both cylindrical and spherical geometry and will be the subject of a future publication.

\section*{Acknowledgements}

The author would like to thank both referees for constructive and helpful comments.
This work has been supported by STFC and by the European Commission through the SOLAIRE Network (MTRN-CT-2006-035484).

\end{document}